# Predicting interviewee attitude and body language from speech descriptors

Yosef Solewicz[1], Chagay Orenshtein[2] and Avital Friedland

[1]National Police, Israel
[2]Tel Hai College, Israel

**Abstract**

This present research investigated the relationship between personal impressions and the acoustic nonverbal communication conveyed by employees being interviewed. First, we investigated the extent to which different conversation topics addressed during the interview induced changes in the interviewees' acoustic parameters. Next, we attempted to predict the observed and self-assessed attitudes and body language of the interviewees based on the acoustic data. The results showed that topicality caused significant deviations in the acoustic parameters statistics, but the ability to predict the personal perceptions of the interviewees based on their acoustic non-verbal communication was relatively independent of topicality, due to the natural redundancy inherent in acoustic attributes. Our findings suggest that joint modeling of speech and visual cues may improve the assessment of interviewee profiles.

# 1. Introduction

**1.1 The study rational**

Human affect sensing can be achieved using a broad range of behavioral cues and signals that are available through diverse channels. Affective states can be recognized based on visible signals, such as gestures (facial expressions, body gestures, head movements, and the like), speech (parameters such as pitch, energy, frequency, and duration), or covert physiological signals (respiratory, cardiac activity, and electrodermal activity).

According to cognitive neuroscience research, information coming from various modalities is combined in our brains to yield multi-modally determined percepts (Driver & Spence, 2000). In real-life situations, our different senses receive correlated information about the same external event. This redundancy can be helpful when some of the channels that convey signals are unavailable, such as in a telephone conversation, when there is no visual feedback from the interlocutor, or in order to enhance speech perception when the audio is corrupted by noise. Furthermore, the multiple signals make it possible for a person assessing someone else's emotional or affective state to consider significantly variable conditions and select alternative channels from the multiple input modalities in order to grasp the emotions being transmitted (Gunes, Piccardi, & Pantic, 2008).

However, what happens if the different information channels send different messages to an observer? People seem to be able to differentiate between honest and untrustworthy channels. In fact, when verbal and nonverbal speech signals contradict each other, people generally trust the latter more, since it unconsciously broadcasts one's true feelings (Ambady & Rosenthal, 1992).

The present research focused on the relationship between visual and acoustic information channels. During interpersonal communication, speech and body gestures coordinate the encoding of nonverbal intents in order to convey underlying internal emotion states (Condon, 1976; Kendon, 1980; McNeill, 1985, 1996). Research has shown that as many as 90% of body gestures are associated with speech, not only regulating the interactions and punctuating discourse, or representing typical social signals but also emphasizing the speakers' thoughts as they occur (see, e.g., McNeill, 1996). These channels are connected at both the behavioral and neural levels (Healey & Braun, 2013). In fact, the relationship between acoustic features and gestures has been the subject of extensive research and some controversy. Although it is widely agreed that gesture and speech reflect the same cognitive process, some researchers have claimed that they are independent and parallel processes (e.g., Krauss, 1998). According to this approach, gestures are seen as an auxiliary channel that supports speech.

In a seminal study, Scherer (2003) presented a theoretical model of the vocal communication of emotion and reviewed the acoustic correlates of different emotional patterns. Cowie et al. (2001) and Juslin and Scherer (2005) provided comprehensive overviews of previous research in the field, and Narayanan and Georgiou (2013) reviewed computational techniques for modeling human behavior based on speech. It is known that emotions can also be visually inferred from gestures, but the mechanisms by which this occurs seem to be unclear (Coulson, 2004). Nevertheless, the combination of speech and visual information has been shown to improve behavior assessment (Busso & Narayanan, 2007; Gatica-Perez, Vinciarelli, & Odobez, 2014; Pantic & Vinciarelli, 2014; Valstar et al., 2013; Yang & Narayanan, 2014).

Unlike the extensive research on speech-emotion mapping published to date, in the present study we did not explicitly address specific emotions or focus on their categorization based on acoustic parameters. Instead, the purpose of the present research was to investigate the general ability to model perceived body language and other expressive intents by means of examining the speaker's acoustic non-verbal communication.

This article follows the sequence of the research process. It begins with analysis of the speech parameters measured in the course of the interviews and demonstration of the significant statistical dissimilarities among the parameters that were extracted from different interview sessions. This is followed by presentation of models for the prediction of visual intents conveyed by interviewees based on the acoustic parameters, using different interview sessions for training and testing. Finally, we report on the results regarding the robustness of the models, which was tested using data from different sessions.

**1.2 Acoustic analysis**

Vocal nonverbal behavior includes five major components: voice acoustics, linguistic and non-linguistic vocalizations, silences, and turn-taking patterns. Each component refers to different social signals that contribute to different aspects of the social perception of a message. In the present research, the acoustic analysis procedure was based entirely on two of these linguistic components: voice acoustics and silences. We estimated the parameters of the acoustics based on stressed vowels only, because they are significantly affected by expressive speech, and in addition, these segments usually possess a high signal-to-noise ratio. In examining the silence excerpts, we focused on discourse pauses, thus considering only relatively long silent segments.

Voice acoustics is a general term, which can be further refined according to the voice production model (for a classical text on this theme, see Rabiner & Schafer, 1978). According to this model, which is often based on the source-filter theory (Fant, 1960), the speech signal is the result of filtering an excitation source. The excitation signal, which is due to airflow from the lungs, passes through the vocal cavity and is shaped into different sounds. For the sake of simplicity, it is assumed that excitation and filter are decoupled, although this is not entirely true. The excitation signal can be roughly classified as voiced or unvoiced. The former is formed by periodic impulses of air modulated by the vocal cords (pitch); in the latter, the excitation is aperiodic.

Based on this model, acoustic features are categorized into three main groups, representing distinct levels within the speech model. The prosody features are those linked to an excitation source at the macro level; they define the intonation and rhythm of speech. At the micro level, the dynamics of the excitation signal define voice-quality characteristics. Finally, spectral (including cepstral) features result from idiosyncrasies of the vocal-tract filter. The features used in the present study and their classification into the three groups – prosodic (P), voice quality (Q), and spectral/cepstral (S) – are presented in Table 1.

Table 1.

*Speech Features Used in the Experiments*

| Feature abbreviation | Type | Feature description |
| --- | --- | --- |
| Spkrate | P | Total vowel length to total speech length ratio |
| Mean pause | P | Mean length of pause segments |
| Pauses second | P | Average number of pauses per second |
| Pause speech ratio | P | Total pause length to total speech length ratio |
| Rhythm | P | Average number of vowels per second |
| Vowel mean | P | Average length of vowels |
| Vowel std | P | Standard deviation of vowel lengths |
| Intensity std | P | Standard deviation of vowel intensity values |
| F0 std | P | F0 standard deviation |
| F0 mean | P | F0 mean |
| Vowel F0 range | P | Average vowel F0 range |
| Harmonicity | Q | Harmonic-to-noise ratio mean |
| Jitter loc | Q | Mean of local jitter |
| Jitter ppq5 | Q | Mean of five-point period perturbation quotient jitter |
| Shimmer loc | Q | Mean of local shimmer |
| Shimmer apq5 | Q | Mean of five-point amplitude perturbation quotient shimmer |
| F1 | S | Mean of first formant frequency |
| F2 | S | Mean of first formant bandwidth |
| F3 | S | Mean of second formant frequency |
| B1 | S | Mean of second formant bandwidth |
| B2 | S | Mean of third formant frequency |
| B3 | S | Mean of third formant bandwidth |
| Cep1 | S | Mean of first mel-freqency cepstral coefficient (MFCC) |
| Cep2 | S | Mean of second MFCC |
| Cep3 | S | Mean of third MFCC |
| Cep4 | S | Mean of forth MFCC |
| Cep5 | S | Mean of fifth MFCC |
| Cep6 | S | Mean of sixth MFCC |
| Cep7 | S | Mean of seventh MFCC |
| Cep8 | S | Mean of eighth MFCC |

## 2. Methodology

The corpus used in this research was formed by a series of recorded interviews in Hebrew with a group of female employees (mean age = 45 years). The interviewees were staff members at daycare centers for infants of low-income families. All the procedures performed in the study that involved human participants were in compliance with the ethical standards of the institutional research committee and with the 1964 Helsinki Declaration and its later amendments or comparable ethical standards. Informed consent was obtained from all the individual participants included in the study.

Two research assistants conducted and recorded the interviews (one in each interview), in secluded rooms at the workplaces. At the end of each interview, the research assistant and the interviewee each completed a questionnaire regarding several aspects of frontal body language, reactions, and their general impression of the interviewee. Both the research assistant and interviewee ranked each attribute on a 7-point scale. The same type of digital recorder device and recording setup were used in all interviews, in order to avoid external distortion of the acoustic features.

During the interview, the employees were given one minute to talk without interruption about a specific theme, in three successive sessions. The aim was to induce different conversation topics, one in each session, during the interview. In the first session, the employee was asked to describe herself in general. In the next session, she was asked to describe a typical workday. In the third and last session, she was asked how she would react to specific hypothetical dangerous situations involving the children with whom she worked.

After filtering out poor recordings, we obtained 297 one-minute recordings (99 speakers x 3 conversation topics). The speech parameters were computed separately for each of these recordings. At this stage, questionnaires that were not properly annotated were also discarded. In the end, we obtained speech data from 69 complete interviews; these served as the basis of the analysis of the body-language attributes defined in the questionnaire.

For the purpose of speech parametrization, the recorded speech files were first converted from mp3 to wav format and then downsampled to 11025 Hz. A phoneme recognizer (Schwarz, Matejka, & Cernocky, 2006) was used to automatically segment each speech file. The vowels detected in all the recognized phonemes were organized in order of length. The longest half of the ordered vowels formed the "stressed vowels" set. A window of 80 ms around the vowel centers was used for prosodic and quality parameter estimation. A shorter window of 40 ms was used for spectral/cepstral parameter estimation. The PRAAT software program (Boersma, 2001) was used for calculating the acoustic features. With regard to discourse pauses, non-voice excerpts longer than 400 ms were considered "pauses" and used to derive the prosodic parameters with temporal characteristics.

As noted earlier, our experiments were divided into two main units: the relationships between several speech parameters and the defined topics, and the prediction of body-language and attitude descriptors based on the speech parameters.

# 3. Experiments

## 3.1 The relationships between speech parameters and conversation topics

*3.1.1 Statistical analysis*

      The initial objective was to determine whether the distinct speech parameters differed significantly by conversation topic. For this purpose, we employed both the paired *t*-test (Goulden, 1956) and the Wilcoxon signed-rank test (Wilcoxon, 1945). The paired *t*-test determines whether two paired sets differ from each other in significantly, based on the assumption that the paired differences are independent and identically normally distributed. The Wilcoxon test is the non-parametric analogue of the paired *t*-test, and should be used if the distribution of differences between pairs is non-normal. In our case, we assessed the differences between speakers in the set of speech features, for pairs of topics. We employed both tests, because some of the features (most notably the temporal features) did not distribute in a typical Gaussian shape. Actually, except in the cases of a few features, both tests yielded the same results. Table 2 shows the test results regarding differences among Topics 1, 2, and 3. For example, $t\ 1\rightarrow 2$ denotes the *t*-test, and W, the Wilcoxon test outcome for a specific feature in the passage from Topic 1 to Topic 2. An upward arrow denotes a positive change in the mean of the feature after moving on to the next topic, and a downward arrow denotes a negative change; blanks reflect no significant change.

Table 2.

*Changes in Speech Parameters in Transition to Conversation Topics* ($N = 99$)

| Feature | $t\ 1\rightarrow 2$ | $t\ 1\rightarrow 3$ | $t\ 2\rightarrow 3$ | $W\ 1\rightarrow 2$ | $W\ 1\rightarrow 3$ | $W\ 2\rightarrow 3$ |
|---|---|---|---|---|---|---|
| Spkrate |  | ↑ | ↑ |  | ↑ | ↑ |
| Mean pause | ↓* |  |  | ↓ | ↓ | ↑ |
| Pauses second | ↓ | ↓ | ↓ | ↓ | ↓ | ↓ |
| Pause speech ratio | ↓ | ↓ |  | ↓ | ↓ |  |
| Rhythm | ↑ | ↑ | ↑ | ↑ | ↑ | ↑ |
| Vowel mean | ↓ | ↓ | ↓ | ↓ | ↓ | ↓ |
| Vowel std |  |  | ↓* |  |  | ↓* |
| Intensity std | ↓* |  | ↑ | ↓* |  | ↑ |
| F0 std |  | ↑* | ↑ |  | ↑* | ↑ |
| F0 mean | ↑ | ↑ | ↑ | ↑ | ↑ | ↑ |

| Parameter | S1 | S2 | S3 | S1 | S2 | S3 |
|---|---|---|---|---|---|---|
| Vowel F0 range | ↓* | ↑* | ↑ | ↓* | ↑* | ↑ |
| Harmonicity | ↑ | | ↓ | ↑ | | ↓ |
| Jitter loc | ↓ | ↑* | ↓ | | | ↑ |
| Jitter ppq5 | ↓* | | | ↓ | | |
| Shimmer loc | ↓ | ↓* | ↑* | ↓ | ↓* | ↑* |
| Shimmer apq5 | ↓ | ↑* | ↓* | | | ↑* |
| F1 | ↑ | ↑ | ↑ | ↑ | ↑ | ↑ |
| F2 | ↓ | ↓ | | ↓ | ↓ | |
| F3 | ↑ | | ↓* | ↑ | | |
| B1 | | ↓ | ↓* | | ↓ | ↓* |
| B2 | | ↓* | | | | |
| B3 | | | | | | |
| Cep1 | | ↓ | ↓ | | ↓ | ↓ |
| Cep2 | ↓ | ↓ | ↓ | ↓ | ↓ | ↓ |
| Cep3 | ↓ | ↓ | ↓ | ↓ | ↓ | ↓ |
| Cep4 | | | | | | |
| Cep5 | | | | ↑* | | |
| Cep6 | ↓ | ↓ | | ↓ | ↓ | |
| Cep7 | | ↑ | ↑* | | ↑ | |
| Cep8 | | | | | | |

\* denotes significance at a level of .05.

*3.1.2 Acoustic correlates*

After the objective statistical analysis of changes among the different speech features across the three induced conversation topics, we proceeded to investigate possible correlations between these attributes and the topics of the different sessions, both quantitative and qualitatively. As noted, during Session 1, the speakers had more freedom regarding choice of the topic; in Session 2, they were directed to focus on work issues, and the third session focused on uncomfortable themes, with the intention of causing the interviewees some degree of stress.

Unlike the majority of previous studies, in the present research we did not direct the interviewees to discuss well-defined topics in order to assess the corresponding speech parameters. Instead, we compared the relative changes in the speech parameters measured across the different topics with those reported in other studies. It should be noted that caution should be exercised in interpreting acoustic correlates of conversation topics, especially in the case of different experiment designs. Different instantiations or variants of specific emotions, even though collectively labeled by the same tag, may be represented by differing acoustic expression patterns (Scherer, 2003).

As Table 2 shows, most of the acoustic parameters did reflect significant differences between the different conversation topics. Taking the interview regarding Topic 1 as the baseline, we investigated the progressive changes in the means of the parameters in Sessions 2 and 3. In general, the spectral/cepstral and most of the prosodic features, shifted towards both topics in the same direction, either positive or negative; this suggests that Topics 2 and 3 were quite similar, and the respective parameters differed from the baseline mostly in the intensity of change. In contrast, the directions of change in the quality descriptors were not consistent among the topics, suggesting that these parameters are more sensitive in capturing qualitative nuances between Topics 2 and 3.

*3.1.3 Quantitative analysis*

To obtain quantitative insights regarding these results, we performed a simplified mathematical analysis. In the following equation, the general acoustic parameter shift from any Topic *a* to Topic *b* is represented as vector *v*. The element *i* of this vector, $v_i$, is the result of a statistical significance test for the $i^{th}$ feature in the shift from Topic *a* to *b*. For simplicity, $v_i$ can be set at one of three values: 1 – rejection of the null hypothesis, $H_0$ – the same statistical distribution for feature *i* in topics *a* and *b* and an increase in the mean for this feature after the move ($\mu_b > \mu_a$); -1 – if it decreases; or 0 – otherwise (Equation 1).

$$v_i = \begin{cases} 1, if\ H_0\ rejected\ and\ \mu_b > \mu_a \\ -1, if\ H_0\ rejected\ and\ \mu_a > \mu_b \\ 0, otherwise \end{cases} \quad (1)$$

Vector ***v*** roughly represents the parameter shift, and therefore, quantification of the overall transition from one conversation topic to another. We can therefore estimate the similarity between two topic transitions by means of vector metrics, such as the Euclidean distance or cosine similarity. This is exemplified using the estimated ***v*** vectors for Topics 1, 2, and 3, and calculating the cosine similarity between pairs of transition topics. Note that a cosine distance between two transitions close to 1 is an indication that the overall acoustic shifts are similar; values close to -1 reflect a negative correlation; and values close to 0 are a sign of uncorrelated changes in the acoustic features.

The numerical results of the scheme described are presented in Table 3. The vectors were estimated using the more general Wilcoxon test, but the *t*-test yielded similar results, as well. Table 3 depicts the cosine similarities among the topic passages for two significance levels, α. It clearly shows that the acoustic changes when the speaker passed from the baseline topic (1) to either Topic 2 or 3 were generally similar. This is expressed by a high $\cos(\vec{12},\vec{13})$) value. On the other hand, the low value found for $\cos(\vec{12},\vec{23})$ suggests that the acoustic changes involved in the transition from Topic 2 to Topic 3 differed from those in the transition from Topic 1 to 2. (Note that since the cosine similarity is not a formal distance metric; it does not have the triangle inequality property.)

Table 3.

*Cosine Similarity Between Distinct Conversation Topics Transitions* ($N = 99$)

| Measurement | Similarity ($\alpha = 0.01$) | Similarity ($\alpha = 0.05$) |
| --- | --- | --- |
| $\cos(\vec{12},\vec{13})$ | .71 | .58 |
| $\cos(\vec{12},\vec{23})$ | .26 | .00 |
| $\cos(\vec{13},\vec{23})$ | .53 | .54 |

However, this process for quantifying similarities between topic transitions lacks further mathematical formality. In particular, it does not consider correlation effects among the different vector components. In future research, a more refined analysis should be conducted on a decorrelated projection of the vector space.

*3.1.4 Qualitative analysis*

We also made a brief attempt to identify traces of emotional speech regarding the different topics. It should be noted that in this research, the changes in speech topicality were not originally meant to lead to specific emotional speech topics. Nevertheless, we found some correlations between the acoustic parameters recorded in the present experiments and the emotional speech patterns reported in the literature (Drioli, Tisato, Cosi, & Tesser, 2003; Grawunder & Winter, 2010; Juslin & Laukka, 2001; Nunes, Coimbra, & Teixeira, 2010; Patel, Scherer, Sundberg, & Bjorkner, 2011; Scherer, 2003; Sobin and Alpert, 1999; Yildirim et al., 2004). It is well accepted that higher-order spectral parameters are generally found to be less sensitive to emotional speech. This was also observed in the present research, due to the absence of statistically significant differences in these parameters among the different topics. On the other hand, mean duration and spread have been found to increase in emotional speech, which was not unequivocally supported in our experiments. F0 mean and spread are expected to increase in positive emotional speech and in fact our results supported this, with the exception of the F0 range of vowels, which decreased in the 1-to-2 passage. This downward pitch inflection might be associated with traces of disgust. Regarding intensity, its spread generally increases in emotional speech (except for sadness). In our experiment, this trend was observed in the passage from Topics 2 to 3, but not in the passage from Topics 1 to 2. Speech rate (including rhythm) usually increases for positive and decreases for negative emotional speech. Our measurements indicated an increase in both positive and negative speech-rate parameters (except for 1-to-2). According to previous studies, F2 tends to increase for emotional speech, and F1 variations depend on the type of emotion. Interestingly, our results showed a decrease in F2 and increase in F1. Finally, voice quality parameters have been found to be an important aspect of emotional speech. Jitter and shimmer seem to be somewhat negative correlated to harmonicity, over distinct emotions. Our results support this general trend. Jitter, shimmer, and pitch variability usually decrease in polite speech. In the present research, this was observed in the passage from Session 1 to 2, but the opposite was found in the passage from Session 2 to 3.

In summary, as expected, the present findings did not indicate clear emotional patterns that characterized the changes between the different interview sessions. Broadly speaking, the 1-2 passage seemed to be emotionally opaque. A possible explanation could be that the task of describing one's work is not really an exciting theme. In comparison, more traces of emotion of the type described in literature were found when the speakers moved to Session 3; this might have been anticipated, considering its sensitive theme. These findings roughly support those of our quantitative analysis, which indicated a more dramatic difference in speech parameters when the speakers moved from Topic 2 to 3 compared with the move from Topic 1 to 2.

**3.2 Predicting body language and attitude from speech parameters**

The second part of our research focused on the ability to model the body language and attitude of interviewees based on assessment of their speech parameters. As reported earlier, the distinct recording sessions within the interviews led to statistically significant differences between the different conversation topics. However, in these experiments, we assumed that the attitudes of the interviewees during the three sessions were not strongly dependent on the specific session and were represented the general perceived (by the research assistants) and self-assessed (by the interviewees) impressions reported at the end of each interview. In other words, we considered the speech parameters indicated for the different topics as independent variables, potentially explaining or predicting the interviewee's behavior traits, the dependent variables. Accordingly, we built separate prediction models for the dependent variables using the assessments of the independent variables for each interview session. The results show that a model trained on given data from a specific session can be used with independent data obtained in other interview sessions; in other words, topicality and conversation topic are relatively irrelevant when training prediction models.

A stepwise linear regression (Draper & Smith, 1998) was employed to create the prediction models. This is an iterative technique for selecting the most statistically significant independent terms to fit a multilinear model for prediction of a dependent variable. One of its limitations is that global optimization is not guaranteed, and different models can be selected under different initial conditions or step sequences. This technique may also suffer from overfitting, which reduces the applicability of models to other datasets. Symptoms of overfitting may be difficult to identify, due to the high rate of correlation among the speech parameters (multicollinearity). A previous decorrelation of the feature space could reduce these problems, and should be considered in the feature.

We built an overall prediction model for each of the dependent variables, but attempted to discard spurious models that could lead to misleading descriptive analyses by means of Leave-One-Out Cross Validation (LOOCV). Specifically, different models were iteratively created for each dependent variable using the left-out sample for testing and the rest of the data for training. We arbitrarily declared a model as stable if:

1. At least 75% of the selected models created during the LOOCV folds were identical (same independent variables selected) to the overall model (trained using the whole data).

2. The ratio between the sample correlation coefficient attained through the LOOCV process and that obtained by the overall model was greater than 0.75.

The following train–test pairs of tables summarize the results obtained for the dependent variables that could be successfully predicted (by means of stable models). The tables present the predicted dependent variable (DV), either as perceived by the interviewer (P) or self-assessed (SA) by the interview; the session (S1, S2, or S3) from which the independent variables were used to train/test the regression models; the selected independent variables (predictors) and their corresponding regression coefficients and correlation signals (positive/negative) in the regression model; and the regression correlation coefficient, *r* (predictors of the same type are placed in a separate line in the tables, for convenience.) Note that we did not perform regression training using SA variables, since they were unavailable for several of the attitude labels.

Table 4a.

*Train Mode for DV "Cooperative"* ($N = 69$)

| Training session | Correlations/predictors | DV type | $R$ |
|---|---|---|---|
| S1 | -.67 Pause–speech ratio -.35 Mean pause +.29 Cep1 | P | .81 |
| S2 | -.73 Pause–speech ratio +.26 B2 | P | .66 |
| S3 | -.78 Pause–speech ratio +.26 Cep6 | P | .70 |

Table 4b.

*Test Mode for DV "Cooperative"* ($N = 69$)

| Trained on | Tested on | DV type | $R$ |
|---|---|---|---|
|    | S2 | P | .59 |
|    | S3 | P | .65 |
| S1 | S1 | SA | .53 |
|    | S2 | SA | .35 |
|    | S3 | SA | .36 |
|    | S1 | P | .74 |
|    | S3 | P | .65 |
| S2 | S1 | SA | .51 |
|    | S2 | SA | .44 |
|    | S3 | SA | .32 |
|    | S1 | P | .73 |
|    | S2 | P | .51 |
| S3 | S1 | SA | .37 |
|    | S2 | SA | .21 |
|    | S3 | SA | .24 |

Table 5a.

*Train Mode for DV "Proposed a practical solution"* ($N = 69$)

| Trained on | Correlations/predictors | DV type | $R$ |
|---|---|---|---|
| S1 | -.77 Pause–speech ratio -.32 Intensity std -.28 Mean pitch | P | .63 |
| S2 | -.71 Pause–speech ratio +.37 B2 | P | .60 |
| S3 | -.77 Pause–speech ratio -.29 Intensity std +.32 Vowel F0 range +.36 Cep6 | P | .70 |

Table 5b.

*Test Mode for DV "Proposed a practical solution"* ($N = 69$)

| Trained on | Tested on | DV type | R |
|---|---|---|---|
| S1 | S2 | P | .56 |
|    | S3 | P | .60 |
|    | S1 | SA | .48 |
|    | S2 | SA | .23 |
|    | S3 | SA | .35 |
| S2 | S1 | P | .54 |
|    | S3 | P | .57 |
|    | S1 | SA | .50 |
|    | S2 | SA | .31 |
|    | S3 | SA | .36 |
| S3 | S1 | P | .54 |
|    | S2 | P | .51 |
|    | S1 | SA | .50 |
|    | S2 | SA | .23 |
|    | S3 | SA | .38 |

Table 6a.

*Train Mode for DV "Serene"* (*N* = 69)

| Trained on | Correlations/predictors | DV type | *R* |
|---|---|---|---|
| S1 | -.83 Pause–speech ratio +.43 cep1 +.60 cep4 | P | .57 |
| S2 | -.71 Pause–speech ratio | P | .42 |
| S3 | -.95 Pause–speech ratio -.54 shimmer apq5 +.67 cep4 +.36 cep6 | P | .70 |

Table 6b.

*Test Mode for "Serene"* (*N* = 69)

| Trained on | Tested on | DV type | *R* |
|---|---|---|---|
| S1 | S2 | P | .41 |
|    | S3 | P | .54 |
|    | S1 | SA | .27 |
|    | S2 | SA | .15 |
|    | S3 | SA | .28 |
| S2 | S1 | P | .43 |
|    | S3 | P | .42 |
|    | S1 | SA | .44 |
|    | S2 | SA | .28 |
|    | S3 | SA | .35 |
| S3 | S1 | P | .46 |
|    | S2 | P | .40 |
|    | S1 | SA | .32 |
|    | S2 | SA | .09 |
|    | S3 | SA | .24 |

Table 7a.

*Train Mode for DV "Hesitant"* (*N* = 69)

| Trained on | Correlations/predictors | DV type | *R* |
|---|---|---|---|
| S1 | +.96 Pause–speech ratio +.40 Vowel std -.56 cep4 | P | .70 |
| S2 | +.65 Pause–speech ratio +.35 Rhythm -.53 cep4 | P | .57 |
| S3 | +.73 Pause–speech ratio +.31 Spkrate -.53 cep4 | P | .60 |

Table 7b.

*Test Mode for DV "Hesitant"* ($N = 69$)

| Trained on | Tested on | DV type | $R$ |
|---|---|---|---|
| S1 | S2 | P | .49 |
|  | S3 | P | .55 |
|  | S1 | SA | .23 |
|  | S2 | SA | .14 |
|  | S3 | SA | .15 |
| S2 | S1 | P | .61 |
|  | S3 | P | .57 |
|  | S1 | SA | .23 |
|  | S2 | SA | .22 |
|  | S3 | SA | .17 |
| S3 | S1 | P | .63 |
|  | S2 | P | .48 |
|  | S1 | SA | .21 |
|  | S2 | SA | .16 |
|  | S3 | SA | .16 |

Table 8a.

*Train Mode for DV "Determined"* ($N = 69$)

| Trained on | Correlations/predictors | DV type | $R$ |
|---|---|---|---|
| S1 | -.96 Pause–speech ratio +.47 Cep1 +.40 Cep4 | P | .66 |
| S2 | -.81 Pause–speech ratio | P | .54 |
| S3 | -.78 Pause–speech ratio +.45 Cep6 | P | .55 |

Table 8b.

*Test Mode for DV "Determined"* ($N = 69$)

| Trained on | Tested on | DV type | $R$ |
|---|---|---|---|
|  | S2 | P | .54 |
|  | S3 | P | .54 |
| S1 | S1 | SA | .38 |
|  | S2 | SA | .22 |
|  | S3 | SA | .38 |
|  | S1 | P | .56 |
|  | S3 | P | .46 |
| S2 | S1 | SA | .44 |
|  | S2 | SA | .33 |
|  | S3 | SA | .41 |
|  | S1 | P | .50 |
|  | S2 | P | .41 |
| S3 | S1 | SA | .31 |
|  | S2 | SA | .17 |
|  | S3 | SA | .37 |

Table 9a.

*Train Mode for DV "Answered properly"* ($N = 69$)

| Trained on | Correlations/predictors | DV type | $R$ |
|---|---|---|---|
| S1 | -.69 Pause–speech ratio -.30 Intensity std +.30 Jitter ppq5 +.37 F2 +.29 Cep1 | P | .64 |
| S2 | -.59 Pause–speech ratio +.44 B2 | P | .59 |
| S3 | -.61 Pause–speech ratio -.26 Intensity std +.30 Vowel F0 range +.41 Cep6 | P | .65 |

Table 9b.

*Test Mode for DV "Answered properly"* ($N = 69$)

| Trained on | Tested on | DV type | $R$ |
|---|---|---|---|
| S1 | S2 | P | .50 |
|  | S3 | P | .57 |
|  | S1 | SA | .23 |
|  | S2 | SA | .16 |
|  | S3 | SA | .08 |
| S2 | S1 | P | .48 |
|  | S3 | P | .49 |
|  | S1 | SA | .18 |
|  | S2 | SA | .09 |
|  | S3 | SA | -.04 |
| S3 | S1 | P | .47 |
|  | S2 | P | .43 |
|  | S1 | SA | .18 |
|  | S2 | SA | .20 |
|  | S3 | SA | .06 |

Table 10a.

*Train Mode for DV "Tremulous"* ($N = 69$)

| Trained on | Correlations/predictors | DV type | $R$ |
|---|---|---|---|
| S1 | +.19 Pause–speech ratio -.26 Cep1 -.40 Cep4 +.22 Cep7 | P | .59 |
| S2 | +.37 Pause–speech ech ratio -.25 B2 +.15 Cep2 -.36 Cep4 | P | .71 |
| S3 | -.34 Mean pause -.26 Pauses second +.66 Pause–speech ratio +.32 Shimmer apq5 -.18 Cep4 -.17 Cep6 | P | .67 |

Table 10b.

*Test Mode for DV "Tremulous"* ($N = 69$)

| Trained on | Tested on | DV type | $R$ |
|---|---|---|---|
| S1 | S2 | P | .53 |
|    | S3 | P | .40 |
| S2 | S1 | P | .36 |
|    | S3 | P | .42 |
| S3 | S1 | P | .22 |
|    | S2 | P | .58 |

Table 11a.

*Train Mode for DV "Turned face aside"* ($N = 69$)

| Trained on | correlations/predictors | DV type | $R$ |
|---|---|---|---|
| S1 | -.41 Jitter loc -.40 Cep1 +.44 Cep7 | P | .47 |
| S2 | +.43 Pause–speech ratio -.56 Cep1 +.42 Cep7 | P | .50 |
| S3 | -.39 Vowel F0 range -.37 B3 -.56 cep6 +.64 Cep7 | P | .60 |

Table 11b.

*Test Mode for DV "Turned face aside"* ($N = 69$)

| Trained on | Tested on | DV type | $R$ |
|---|---|---|---|
| S1 | S2 | P | .40 |
|    | S3 | P | .43 |
| S2 | S1 | P | .35 |
|    | S3 | P | .21 |
| S3 | S1 | P | .30 |
|    | S2 | P | .26 |

Table 12a.

*Train Mode for DV "Breathed rapidly"* (*N* = 69)

| Trained on | Correlations/predictors | DV type | *R* |
|---|---|---|---|
| S1 | +.46 Mean pause -.56 Cep4 | P | .65 |
| S2 | -.31 Pauses second +.65 Pause–speech ratio +.22 F2 -.26 Cep4 | P | .71 |
| S3 | -.34 Pauses second +.74 Pause–speech ratio +.44 Shimmer apq5 -.42 Cep4 | P | .74 |

Table 12b.

*Test Mode for DV "Breathed rapidly"* (*N* = 69)

| Trained on | Tested on | DV type | *R* |
|---|---|---|---|
| S1 | S2 | P | .55 |
| S1 | S3 | P | .50 |
| S2 | S1 | P | .58 |
| S2 | S3 | P | .65 |
| S3 | S1 | P | .58 |
| S3 | S2 | P | .56 |

## 4. Discussion

This research investigated the ability to predict body language and behavioral traits based on speech descriptors during interviews. The body language and behavioral reactions used in the models were either collected as perceived by two research assistants and self-assessment opinions. The former option led to better modeling, which could be attributed to the more consistent ranking scale used by the assistants. Subjects were recorded on three distinct sessions. In general, in all three sessions, stable prediction models could be trained successfully. Examples of less successful models were those attempting to predict: eye/mouth/lip/hand/finger movements, posture, coughing, scratching, laughing, joyfulness, blushing, and stress.

One of the interesting findings of this research concerns the general freedom allowed regarding discourse topicality in the creation of the prediction models. As shown earlier, different interview sessions focusing on distinct topics differed significantly in terms of several speech descriptors and also led to different prediction models. Nevertheless, our results show that equally efficient models could be trained and further tested on speech parameters processed from different interview sessions. This further support the idea that speech information generally represents a flexible, well-synchronized, and robust channel for decoding the visual intents conveyed by the interviewee.

Regarding the composition of the prediction models, a combination of prosodic, spectral, and voice-quality predictors was often found. According to the regressions, prosodic predictors, and in particular *pause speech ratio* seemed to be the predominant predictors. Furthermore, this feature also had the highest relative coefficient weight in the regression models. On the other hand, voice-quality features emerged as the least relevant.

More specifically, all the predictors included a combination of *pause–speech ratio* and some cepstral parameters. In most cases, positive reactions (cooperative, practical, serene, determined, answered properly) were characterized by a decrease in *pause–speech ratio* (consistent with previous findings of Baskett & Freedle, 1974 and Scherer, 1979 that long pauses and interlocutor latency induce the perception of incompetence) accompanied by an increase in cepstral parameters, and the opposite was true with regard to negative reactions (hesitant, tremulous, turned face aside, breathing). (Note that cep 7 consistently displayed an opposite trend compared with other cepstral coefficients; see Table 2.)

Following is a brief summary of other interesting prediction models obtained in the research. Hesitation was characterized by an increase in speech rate, rhythm and variation in vowel length, accompanied by increased periods of silence, which indicate speech in bursts (Although rapid speech rates have also been associated with competence and sociability (Miller, Maruyama, Beaber, & Valone, 1976), the sensation of hesitance was probably caused by the increase in silences.)

A combination of increases in the *vowel F0 range* accompanied by a decrease in *intensity std* was selected in two positive reactions (practical, answered properly). According to Hirschberg, 1993, an increase in pitch within certain words is used in an effort to structure the discourse. A decrease in the *vowel F0 range* was spotted in a negative reaction (face aside).

The less prevalent voice quality parameters eventually played an important role in discriminating between excited (breathed rapidly, tremulous, answered properly) and opaque reactions (face aside, serene). Excitation was accompanied by an increase in shimmer and jitter, as also observed by Chung (2000). Decreasing trends in these parameters indicated opaque reactions and show agreement with similar measurements found for polite speaking modes (Grawunder & Winter, 2010).

One of the strengths of the present study was the homogeneity of the research population. All the participants belonged to the same organization and shared a similar social background. We also enriched the data analysis by employing distinct speech topics and using both external- and self-assessed subjective interviewee evaluations. In addition, we used the same recording device and setup during all the recordings. Finally, only two research assistants conducted the recordings and the subjective evaluations. All these factors contributed to the consistency in the results by reducing the amount of noise on the measured parameters, at the level of both contents and processing.

A few weaknesses should also be noted. These include some ambient noise and background speech in the recordings and the limited scope of the research population, which included only women and was relatively small. We also note the limitations regarding different artificially induced pressure scenarios during the interviews.

These shortcomings notwithstanding, the present research contributes an additional step forward in the understanding of body language, by examining its relationship with visual and auditory variables in interview situations. A direct application of the reported results could be the development of protocols for analyzing or profiling interviewee behavior during audio or video chats, in particular, regarding Performance Appraisal Interviews (PAI) (Asmuß (2013) and Asmuß (2008)). This kind of interview discusses the performance of an employee vis-à-vis with his employer and involves scenarios relatively similar to our experimental setup. In further studies, we plan to study additional varied research populations, different scenarios, and additional non-verbal communication variables.